\documentclass[a4paper,5p]{elsarticle}
\usepackage{graphicx,dcolumn,bm,amssymb,amsmath,hyperref,color,natbib}
\usepackage{tgtermes}

\newcommand{\ud}{\mathrm{d}}
\newcommand{\ve}{\varepsilon}

\journal{Physics Letters B}

\biboptions{sort&compress}

\begin{document}
\begin{frontmatter}

\title{What $\xi$ ?
Cosmological constraints on the non-minimal coupling constant}

\author{Orest Hrycyna}
\address{Theoretical Physics Division, National Centre for Nuclear
Research, Ho{\.z}a 69, 00-681 Warszawa, Poland}

\ead{orest.hrycyna@ncbj.gov.pl}

\begin{abstract}
In dynamical system describing evolution of universe with the flat Friedmann-Robertson-Walker symmetry filled with barotropic dust matter and non-minimally coupled scalar field with a constant potential function an invariant manifold of the de Sitter state is used to obtain exact solutions of the reduced dynamics. Using observational data coming from distant supernovae type Ia, the Hubble function $H(z)$ measurements and information coming from the Alcock-Paczy{\'n}ski test we find cosmological constraints on the non-minimal coupling constant $\xi$ between the scalar curvature and the scalar field. For all investigated models we can exclude negative values of this parameter at the $68\%$ confidence level. We obtain constraints on the non-minimal coupling constant consistent with condition for conformal coupling of the scalar field in higher dimensional theories of gravity.

\end{abstract}


\end{frontmatter}

\section{Introduction}

The discovery of accelerated expansion of the universe \cite{Riess:1998cb,Perlmutter:1998np} drastically changed our understanding of the history and future of our universe. The simplest candidate for the dark energy driving the current phase of the evolution of the universe seems to be a positive cosmological constant and the straightforward application of this assumption within the general theory of relativity together with the Friedmann-Robertson-Walker symmetry lead to the $\Lambda$CDM model which is a standard model of the cosmological evolution \cite{Ade:2013zuv,Ade:2015xua} and is favoured by the observational data \cite{Kurek:2007tb,Kurek:2007gr,Szydlowski:2006ay}. However such a explanation suffers from the fine tuning problem \cite{Weinberg:1988cp, Sahni:1999gb} and the coincidence problem \cite{Sahni:1999gb, Zlatev:1998tr}. These problems stimulated investigations in the field of dynamical dark energy models \cite{Copeland:2006wr}. The quintessence idea was formulated \cite{Ratra:1987rm, Wetterich:1987fm} as a simplest model involving a scalar field with a potential function in order to describe the current accelerated expansion of the universe. Soon after number of alternatives were proposed in order to alleviate and eliminate problems with a cosmological constant term like a phantom dark energy \cite{Caldwell:2003vq,Dabrowski:2003jm} or a extended quintessence \cite{Perrotta:1999am, Faraoni:2000gx, Faraoni:2006ik, Hrycyna:2007mq, Hrycyna:2007gd, Hrycyna:2008gk, Hrycyna:2009zj, Hrycyna:2010yv}.

The simplest generalisation of the scalar field Lagrangian is the inclusion into a matter sector of the theory a non-minimal coupling term between the gravity and the scalar field of the form $-\xi R\phi^{2}$ \cite{Chernikov:1968zm, Callan:1970ze, Birrell:1979ip}. The motivations for this term can be found in different contexts. The general 
relativity has a methodological status of an effective theory with a given Lagrangian and such 
contribution naturally emerge in its expansion \cite{Donoghue:1994dn}. The 
non-minimal coupling between the scalar curvature and the scalar field appears as a result of quantum corrections to the scalar field in curved space \cite{Allen:1983dg, Ishikawa:1983kz, Birrell:1984ix, Parker:book} and is required by the renormalisation procedure \cite{Callan:1970ze}. The 
non-minimal coupling is also interesting in the context of 
superstring theory \cite{Maeda:1985bq} and induced gravity \cite{Accetta:1985du}. While the simplest inflationary model with a minimally coupled scalar field and a quadratic potential function is no longer favoured by the observational data \cite{Planck:2013jfk, Martin:2013nzq, Ade:2015lrj, Kobayashi:2011nu} there is a need to extend this paradigm further. 
From the theoretical point of view and an effective theory approach the coupling constant becomes a free parameter in the model and should be obtained from some general considerations \cite{Atkins:2010eq, Atkins:2010re} or from a more fundamental theory.
Taking a pragmatic approach its value should be estimated from the observational data \cite{Luo:2005ra, Nozari:2007eq, Szydlowski:2008zza, Atkins:2012yn}.

The non-minimally coupled scalar field cosmology was investigated by many authors in the connection with an inflationary epoch as well as a description of the current accelerated expansion of the universe \cite{Spokoiny:1984bd, Belinsky:1985zd, Ford:1986sy, Salopek:1988qh, Amendola:1990nn, Demianski:1991zv,Fakir:1992cg, Barvinsky:1994hx, Faraoni:1996rf, Barvinsky:1998rn, Barvinsky:2008ia, Setare:2008mb, Setare:2008pc, Uzan:1999ch, Chiba:1999wt, Amendola:1999qq, Holden:1999hm, Bartolo:1999sq, Boisseau:2000pr, Gannouji:2006jm, Carloni:2007eu, Bezrukov:2007ep, Kamenshchik:1995ib}. In the standard model of particle physics a non-minimally coupled Higgs field plays also important role \cite{DeSimone:2008ei, Bezrukov:2008ej, Barvinsky:2009fy, Clark:2009dc}.

The dynamical systems methods are widely used in cosmological applications since seminal papers by Belinskii \cite{Belinskii:1985, Belinskii:1987} and the most widespread parameterisation of the phase space is by the so-called expansion normalised variables \cite{Copeland:1997et} (see collection of works on the dynamical system analysis of anisotropic models \cite{Wainwright:book}). In the present paper we find cosmological constraints on the non-minimal coupling constant starting from background dynamics of a simple cosmological model with a constant potential function. Using phase space variables normalised in the present epoch we show that dynamical system describing evolution of the model is equipped with an invariant manifold corresponding to the de Sitter type of expansion. This manifold enables us to reduce dynamics and we find exact solutions of the reduced dynamical system.  Using observational data from the recent history of the universe we obtain constraints on the non-minimal coupling constant consistent with condition for conformal coupling of the scalar field in higher dimensional theories of gravity.

\section{The model and the method}

We start from the total action of the theory 
\begin{equation}
S =S_{g} + S_{\phi}+ S_{m}\,,
\end{equation}
consisting of the gravitational part described by the standard Einstein-Hilbert action integral
\begin{equation}
\label{eq:gravity}
S_{g} = \frac{1}{2\kappa^{2}}\int\ud^{4}x\sqrt{-g}\,R \,,
\end{equation}
where $\kappa^{2}=8\pi G$, and the matter part of the theory is composed of two substances. One is in the form of non-minimally coupled scalar field
\begin{equation}
\label{eq:scalar}
S_{\phi}=- \frac{1}{2}\int\ud^{4}x\sqrt{-g}\Big(\ve\nabla^{\alpha}\phi\,\nabla_{\alpha}\phi + \ve\xi R 
\phi^{2} + 2U(\phi)\Big)\,,
\end{equation}
where $\ve=+1,-1$ corresponds to the canonical and the phantom scalar field, respectively, and the second one in the form of barotropic matter
\begin{equation}
S_{m} = \int\ud^{4}x\sqrt{-g}\mathcal{L}_{m}\,.
\end{equation}

The field equations for the theory are
\begin{equation}
R_{\mu\nu}-\frac{1}{2}g_{\mu\nu}R=\kappa^{2}\left(T^{(\phi)}_{\mu\nu}+T^{(m)}_{\mu\nu}\right)\,,
\end{equation}
where the energy-momentum tensor for the non-minimally coupled scalar field is given by
\begin{equation}
\label{eq:energy_mom}
\begin{split}
T^{(\phi)}_{\mu\nu}= \,\,& \ve \nabla_{\mu}\phi\nabla_{\nu}\phi -
\ve\frac{1}{2}g_{\mu\nu}\nabla^{\alpha}\phi\nabla_{\alpha}\phi -
U(\phi)g_{\mu\nu} + \\ & + \ve\xi\Big(R_{\mu\nu}-\frac{1}{2}g_{\mu\nu}R\Big)\phi^{2} +
\ve\xi\Big(g_{\mu\nu}\Box\phi^{2}-\nabla_{\mu}\nabla_{\nu}\phi^{2}\Big)\,,
\end{split}
\end{equation}
and for the barotropic matter 
\begin{equation}
T^{(m)}_{\mu\nu}= \big(\rho_{m}+p_{m}\big)u_{\mu}u_{\nu} + p_{m}g_{\mu\nu}\,.
\end{equation}
Finally, from the variation $\delta S_{\phi}/\delta\phi=0$ we obtain the dynamical equation for the scalar field
\begin{equation}
\Box\phi-\xi R\phi-\ve \,U'(\phi)=0\,.
\end{equation}

The non-minimal coupling between the scalar field and the curvature leads to important fact that the field equations can be written in several nonequivalent ways. In the present paper the energy momentum tensor for the scalar field \eqref{eq:energy_mom} is covariantly conserved, which may not be true for other possibilities \cite{Faraoni:2000wk, Faraoni:book, Capozziello:book}. On the other hand in cosmological applications one can obtain substantial mathematical simplification using conformal transformation techniques, especially in the absence of ordinary matter. This procedure enables one to relate cosmological models with a non-minimally coupled scalar field with its conformal counterpart with a minimally coupled field. The Jordan frame action integral, where the scalar field is non-minimally coupled to the Ricci scalar curvature is mapped into an Einstein frame where now the transformed scalar field $\tilde{\phi}$ is minimally coupled. The two frames are physically nonequivalent unless variable units of time, length, and mass are adopted in the Einstein frame \cite{Faraoni:book, Capozziello:book}.


Here we work exclusively in the Jordan frame formulation of the theory leaving aside the question concerning equivalence between the Jordan frame and the Einstein frame formulation of gravitational theory \cite{Faraoni:1999hp, Kamenshchik:2014waa}.

From now on we assume that the geometry is given by the flat Friedmann-Robertson-Walker metric
\begin{equation}
\ud s^{2} = -\ud t^{2} +a^{2}(t)\Big(\ud x^{2}+\ud y^{2}+\ud z^{2}\Big)\,,
\end{equation}
and that the barotropic matter is in form of a dust matter.

The energy conservation condition is in the standard form
\begin{equation}
\label{eq:en_con}
\frac{3}{\kappa^{2}}H^{2} = \rho_{\phi} + \rho_{m}
\end{equation}
where the energy density of the scalar field with a constant potential function $U(\phi)=U_{0}=\text{const.}$ is 
\begin{equation}
\rho_{\phi} = \ve\frac{1}{2}\dot{\phi}^{2}+ U_{0} + \ve3\xi H^{2}\phi^{2} +
\ve3\xi H (\phi^{2})\dot\,,
\end{equation}
and a dot denotes differentiation with respect to cosmic time $t$ and $\rho_{m}$ is the energy density of the barotropic dust matter. The trace of the field equations give the acceleration equation in the following form
\begin{equation}
\label{eq:accel}
\dot{H}=-2H^{2}+\frac{\kappa^{2}}{6}\frac{-\ve(1-6\xi)\dot{\phi}^{2}+4U_{0}+\rho_{m}}{1-\ve\xi(1-6\xi)\kappa^{2}\phi^{2}}\,,
\end{equation}
and the dynamics of the scalar field is governed by the equation
\begin{equation}
\label{eq:dyn_phi}
\ddot{\phi}+3H\dot{\phi}+6\xi\phi(\dot{H}+2H^{2})=0\,.
\end{equation}
Dynamical properties of the model are completely described by the system of equations \eqref{eq:accel} and \eqref{eq:dyn_phi} subject to the energy conservation condition \eqref{eq:en_con}. This is why we can reduce dynamics to a simple autonomous dynamical system introducing the following phase space variables \cite{Hrycyna:2015eta}
\begin{equation}
x\equiv\frac{\kappa\dot{\phi}}{\sqrt{6}H_{0}}\,,\quad z\equiv\frac{\kappa}{\sqrt{6}}\frac{H}{H_{0}}\phi\,,\quad h\equiv\frac{H}{H_{0}}\,,
\end{equation}
together with the dimensionless energy density parameters
\begin{equation}
\Omega_{m}\equiv \frac{\kappa^{2}\rho_{m}}{3 H_{0}^{2}}\,,\quad \Omega_{\Lambda,0} \equiv \frac{\kappa^{2}U_{0}}{3 H_{0}^{2}}\,.
\end{equation}
The energy conservation condition, a modified Friedmann equation, is 
\begin{equation}
\label{eq:constr}
\begin{split}
& \left(\frac{H(a)}{H(a_{0})}\right)^{2} = h^{2} = \\ & = \Omega_{\Lambda,0}+
\Omega_{m,0}\left(\frac{a}{a_{0}}\right)^{-3} +
\ve(1-6\xi)x^{2} + \ve6\xi(x+z)^{2}\,,
\end{split}
\end{equation}
and the acceleration equation now is given by
\begin{equation}
\frac{\dot{H}}{H^{2}} = -2 +
\frac{-\ve(1-6\xi)x^{2}+2\Omega_{\Lambda,0} +
\frac{1}{2}\Omega_{m,0}\left(\frac{a}{a_{0}}\right)^{-3}}
{h^{2}-\ve6\xi(1-6\xi)z^{2}}\,,
\end{equation}
where we can eliminate the $\Omega_{m,0}$ term using the energy conservation condition
\eqref{eq:constr}.

Finally, the dynamical system on variables $x$, $z$ and $h$ is in the
following form
\begin{equation}
\label{eq:dynsys1}
\begin{split}
\frac{\ud x}{\ud \ln{a}} &= -3x-6\xi z\Big(\frac{\dot{H}}{H^{2}}+2\Big)\,,\\
\frac{\ud z}{\ud\ln{a}} &= x+z\frac{\dot{H}}{H^{2}}\,, \\
\frac{\ud h}{\ud\ln{a}} &= h \frac{\dot{H}}{H^{2}}\,,
\end{split}
\end{equation}
where
\begin{equation}
	\frac{\dot{H}}{H^{2}} = -2
	+\frac{3}{2}\frac{\Omega_{\Lambda,0}+\frac{1}{3}h^{2}-
	\ve(1-6\xi)x^{2}-\ve2\xi(x+z)^{2}}{h^{2}-\ve6\xi(1-6\xi)z^{2}} \,.
\end{equation}
From the last equation of the dynamical system \eqref{eq:dynsys1}, one can notice, that the system has two invariant manifolds
$h=H/H_{0}=0$ and $\dot{H}/H^{2} =0$. Especially the latter one is the most interesting from the physical point of view since it corresponds to the de Sitter type of evolution. In theory of dynamical systems we are interested in asymptotic sates which in cosmological applications correspond to different phases of evolution of the universe. Such asymptotic states exist in two forms, the critical points for which the right hand sides of the dynamical system vanish identically and the invariant manifolds for which one of the dynamical equations vanishes  \cite{Wiggins:book_im,Perko:book,Wiggins:book}. If during the evolution the system hits the invariant manifold it stays there forever and the dimensionality of the dynamical system reduces. The simplest example of an invariant manifold in dynamical system is, of course, an arbitrary phase curve which in a special case can be in form of a separatrix of a saddle type critical point. In cosmological applications of dynamical system theory where we usually use the scale factor as a time parameter along phase curves a vacuum model or a flat FRW model constitute an invariant manifold when investigating the dynamics of models with matter or closed (open) FRW models.  

On the invariant manifold $\frac{\dot{H}}{H^{2}}=0$, corresponding to the de Sitter type of evolution, the system \eqref{eq:dynsys1} reduces to the following linear system
\begin{equation}
\label{eq:dynsys_red}
\begin{split}
	\frac{\ud x}{\ud \ln{a}} & = -3x -12\xi z\,,\\
	\frac{\ud z}{\ud \ln{a}} & = x\,,
\end{split}
\end{equation}
which exact solutions can be easily found and the initial conditions for the phase space variables $x_{0}$, $z_{0}$ and the scale factor $a_{0}$ can be arbitrary chosen and we choose them at the present epoch.

Using the exact solutions on the invariant manifold one can try to solve the third differential equation in the system \eqref{eq:dynsys1} in order to obtain the Hubble function governing the background expansion of the universe. In our previous paper \cite{Hrycyna:2015eta}, using appropriate Taylor series expansion in one of the phase space variable, we were able to show that these linear solutions naturally lead to the Hubble function where the dominant terms correspond to the standard $\Lambda$CDM model and additional terms which constitute extensions to the standard cosmological model crucially depend on the value of non-minimal coupling constant $\xi$. Additionally, the quadratic terms in initial conditions $(x_{0},z_{0})$ vanish for $\xi=\frac{3}{16}$. 

In the present paper we want to go much further in order to obtain the observational constraints on the non-minimal coupling constant $\xi$. Our starting point are the exact solutions of the reduced system \eqref{eq:dynsys_red} and then we numerically integrate the third equation of the system \eqref{eq:dynsys1}
\begin{equation}
\label{eq:accel_int}
\begin{split}
& \frac{1}{h(a)}\frac{\ud h(a)}{\ud\ln{a}} 
= \\ & \left(-2
	+\frac{3}{2}\frac{\Omega_{\Lambda,0}+\frac{1}{3}h^{2}(a)-
	\ve(1-6\xi)x^{2}(a)-\ve2\xi\big(x(a)+z(a)\big)^{2}}{h^{2}(a)-\ve6\xi(1-6\xi)z^{2}(a)}\right)\,,
	\end{split}
\end{equation}
with the initial condition for the Hubble function at the present epoch 
\begin{equation}
h(a_{0})=h_{0}=1\,.
\end{equation}
Using the energy conservation condition \eqref{eq:constr} one can further constrain the value of the energy density associated with value of the potential function
\begin{equation}
\Omega_{\Lambda,0}=1-\Omega_{m,0}-\ve(1-6\xi)x_{0}^{2} - \ve6\xi(x_{0}+z_{0})^{2}\,.
\end{equation}
In the result we obtain the numerically calculated Hubble function
\begin{equation}
\frac{H(a)}{H(a_{0})}=h(a,\Omega_{m,0},\ve,\xi,x_{0},z_{0})\,,
\end{equation}
for a given set of parameters of the model $(\Omega_{m,0},\ve,\xi,x_{0},z_{0})$. First we distinguish between possible scalar fields, canonical $\ve=+1$ or phantom $\ve=-1$, next we assume two possibilities of the barotropic matter content. One is represented by the pure baryonic matter $\Omega_{m,0}=\Omega_{bm,0}$ and the next one is equipped with additional contribution from the dark matter $\Omega_{m,0}=\Omega_{bm,0}+\Omega_{dm,0}$. The last thing is to constrain the value of $\Omega_{\Lambda,0}$ parameter. In this way we obtain eight possible different cosmological models with different matter content.

Note that for the minimally coupled scalar field with $\xi=0$ we obtain the following Hubble function
\begin{equation}
	\label{eq:Hmin}
	h^{2}(a)\big|_{\xi=0}=
	\Omega_{\Lambda,0} + \Omega_{m,0}\left(\frac{a}{a_{0}}\right)^{-3} + \ve\,
	x_{0}^{2}\left(\frac{a}{a_{0}}\right)^{-6}\,,
\end{equation}
where $\Omega_{\Lambda,0}=1-\Omega_{m,0}-\ve x_{0}^{2}$. This formula describes the standard $\Lambda$CDM model where the last term indicates contribution from the scalar field and has a similar form as a stiff matter. Depending on the type of scalar field (canonical $\ve=+1$ or phantom $\ve=-1$) this contribution has positive or negative energy density. The observational constraints on such modified Hubble function lead to conclusion that this model is indistinguishable from the $\Lambda$CDM model \cite{Szydlowski:2008zz}.

Within the assumption about conformal coupling of the scalar field with $\xi=\frac{1}{6}$ we obtain
\begin{equation}
	\label{eq:Hconf}
	h^{2}(a)\big|_{\xi=\frac{1}{6}}=
	\Omega_{\Lambda,0} + \Omega_{m,0}\left(\frac{a}{a_{0}}\right)^{-3} + \ve\,
	(x_{0}+z_{0})^{2}\left(\frac{a}{a_{0}}\right)^{-4}\,,
\end{equation}
where $\Omega_{\Lambda,0}=1-\Omega_{m,0}-\ve (x_{0}+z_{0})^{2}$.
The resulting Hubble function describes the standard $\Lambda$CDM model with additional term coming from the conformally coupled scalar field which behaves like a radiation in the model. The energy density of this radiation-like term is positive for a canonical scalar field and is negative for a phantom scalar field. Within a model with conformal coupling and containing dark radiation a simple bouncing solution tending to de Sitter space can be found \cite{Boisseau:2015hqa}.

Finally, for the special initial conditions $x_{0}=z_{0}=0$ from equation \eqref{eq:accel_int} we obtain
\begin{equation}
h^{2}(a)=\left(\frac{H(a)}{H(a_{0})}\right)^{2}=\Omega_{\Lambda,0}+\Omega_{m,0}\left(\frac{a}{a_{0}}\right)^{-3}\,,
\end{equation}
which represents the $\Lambda$CDM model. In this sense we investigate simple extensions of the standard cosmological model caused by the non-minimal coupling between the scalar curvature and the scalar field.

\section{Observational constraints}

Estimations of the parameters of the model were made using, modified for our purposes, publicly available \textsc{CosmoMC} source code \cite{cosmomc,Lewis:2002ah} with implemented nested sampling algorithm \textsc{multinest} \cite{Feroz:2007kg,Feroz:2008xx,Feroz:2013hea} together with explicit Runge-Kutta method of order $8$ with dense output of order $7$ \cite{Hairer:codes,Hairer:book} for numerical solutions of the acceleration equation. The present values of the Hubble function $H_{0}= 67.27\,\, \text{Mpc/km/s}$ and the baryonic matter density parameter $\Omega_{bm,0}\text{h}^{2} = 0.02225$ taken from the observations of the Planck satellite \cite{Ade:2015xua} were kept fixed. In all models under consideration we assumed a flat prior for estimated parameters in the following intervals: $x_{0}\in(-5;5)$, $z_{0}\in(-5;5)$, $\xi\in(-1;4)$ and $\Omega_{dm,0}\in(0;1)$.

In the approach presented in this paper the starting point was the numerically derived Hubble function \eqref{eq:accel_int} together with the exact solutions of the phase space variables on the invariant de Sitter manifold. The initial conditions for the phase space variables $x_{0}$ and $z_{0}$ together with parameter of the theory $\xi$ occurring in \eqref{eq:accel_int} are treated on equal footing and a flat prior was assumed since they are parameters of the Hubble function and there is no prior knowledge about values of those parameters. The interval for the $\xi$ parameter was imposed by dynamical system analysis of the model where $\xi$ was treated as a bifurcation parameter leading to identification of bifurcation values giving rise to qualitatively different evolutional paths of the universe \cite{Hrycyna:2010yv, Hrycyna:2015eta}. 

During the parameters estimations we used the observational data of 580 supernovae type Ia, the so called \textsc{Union2.1} compilation \cite{Suzuki:2011hu}, 31 observational data points of Hubble function from \cite{Jimenez:2001gg,Simon:2004tf,Gaztanaga:2008xz,Stern:2009ep,Moresco:2012jh,Busca:2012bu,Zhang:2012mp,Blake:2012pj,Chuang:2012qt,Anderson:2013oza} collected in \cite{Chen:2013vea} and information coming from determinations of Hubble function using Alcock-Paczy\'{n}ski test \cite{Alcock:1979mp,Blake:2011ep}.

The likelihood function for the supernovae data is defined by
\begin{equation}
L_{SN} \propto \exp \left[ - \sum_{i,j}(\mu_{i}^{\mathrm{obs}} - \mu_{i}^{\mathrm{th}}) \mathbb{C}_{ij}^{-1} (\mu_{j}^{\mathrm{obs}} - \mu_{j}^{\mathrm{th}})\right] \, ,\label{sn_likelihood}
\end{equation}
where $\mathbb{C}_{ij}$ is the covariance matrix with the systematic errors, $\mu_{i}^{\mathrm{obs}}=m_{i}-M$ is the distance modulus, $\mu_{i}^{\mathrm{th}}=5\log_{10}D_{Li} + \mathcal{M}=5\log_{10}d_{Li} + 25$, $\mathcal{M}=-5\log_{10}H_{0}+25$ and $D_{Li}=H_{0}d_{Li}$, where $d_{Li}$ is the luminosity distance which is given by $d_{Li}=(1+z_{i})c\int_{0}^{z_{i}} \frac{dz'}{H(z')}$ (with the assumption $k=0$).

For $H(z)$ the likelihood function is given by
\begin{equation}
L_{H(z)} \propto \exp \left[ - \sum_i\frac{\left(H^{\mathrm{th}}(z_i)-H^{\mathrm{obs}}_i\right)^2}{2 \sigma_i^2} \right ],
\label{hz_likelihood}
\end{equation}
where $H^{\mathrm{th}}(z_i)$ denotes the theoretically estimated Hubble function, $H^{\mathrm{obs}}_i$ is observational data.

And finally, the likelihood function for the information coming from Alcock-Paczy\'{n}ski test is given by
\begin{equation}
L_{AP} \propto \exp \left[ - \sum_i\frac{\left(AP^{\mathrm{th}}(z_i)-AP^{\mathrm{obs}}_i\right)^2}{2 \sigma_i^2} \right ],
\label{ap_likelihood}
\end{equation}
where: $AP^{\mathrm{th}}(z_i)\equiv \frac{H(z_i)}{H_0 (1+z_i)}$.

The total likelihood function $L_{TOT}$ is defined as
\begin{equation}
L_{TOT}=L_{SN}L_{H(z)}L_{AP}.
\label{total_likelihood}
\end{equation}

\begin{table*}
\renewcommand{\arraystretch}{1.5}
	\centering
	\begin{tabular}{lccl}
		\hline
model & $\xi$ & $\Omega_{dm,0}$ & assumptions \\
		\hline
canonical $1$	
& $0.1837^{+0.0513}_{-0.0598}$& -- &$\Omega_{m,0}=\Omega_{bm,0}$ \\
phantom $1$	
& $0.2449^{+0.0694}_{-0.0624}$ & -- &$\Omega_{m,0}=\Omega_{bm,0}$ \\
canonical $2$	
& $0.2180^{+0.0329}_{-0.0257}$& -- & $\Omega_{m,0}=\Omega_{bm,0}$ , $\Omega_{\Lambda,0}>0$\\
phantom $2$	 
& $0.2474^{+0.0633}_{-0.0563}$& -- & $\Omega_{m,0}=\Omega_{bm,0}$ , $\Omega_{\Lambda,0}>0$\\
		\hline
canonical $1+$dm	
& $0.2480^{+0.1158}_{-0.1181}$ & $0.2875^{+0.0891}_{-0.1032}$ & $\Omega_{m,0}=\Omega_{bm,0}+\Omega_{dm,0}$ \\
phantom $1+$dm	
& $0.2203^{+0.0800}_{-0.1293}$ & $0.3032^{+0.0945}_{-0.0995}$ &$\Omega_{m,0}=\Omega_{bm,0}+\Omega_{dm,0}$ \\
canonical $2+$dm	 
& $0.4967^{+0.3496}_{-0.3303}$& $0.2873^{+0.0689}_{-0.0903}$& $\Omega_{m,0}=\Omega_{bm,0}+\Omega_{dm,0}$ , $\Omega_{\Lambda,0}>0$\\
phantom $2+$dm	
& $0.2319^{+0.0970}_{-0.1557}$& $0.3045^{+0.0985}_{-0.1021}$& $\Omega_{m,0}=\Omega_{bm,0}+\Omega_{dm,0}$ , $\Omega_{\Lambda,0}>0$\\
\hline
	\end{tabular}
	\caption{Mean of marginalised posterior probability distribution functions with $68\%$ confidence levels for the non-minimal coupling constant $\xi$ and the barotropic dark matter density parameter of the investigated models.}
\label{tab:1}
\end{table*}

The mean of marginalised posterior probability distribution functions with $68\%$ confidence level of the non-minimal coupling constant $\xi$ and the barotropic dark matter density parameter for all investigated models are gathered in table \ref{tab:1}. In the first part we presented canonical $\ve=+1$ and phantom $\ve=-1$ models without barotropic dark matter, while in the second part of the table we presented results for the same models but with additional contribution in the form of the barotropic dark matter. For all investigated models we obtain the positive values of the non-minimal coupling constant $\xi$ at the $68\%$ confidence level.

\begin{figure*}
\centering
\includegraphics[scale=0.425]{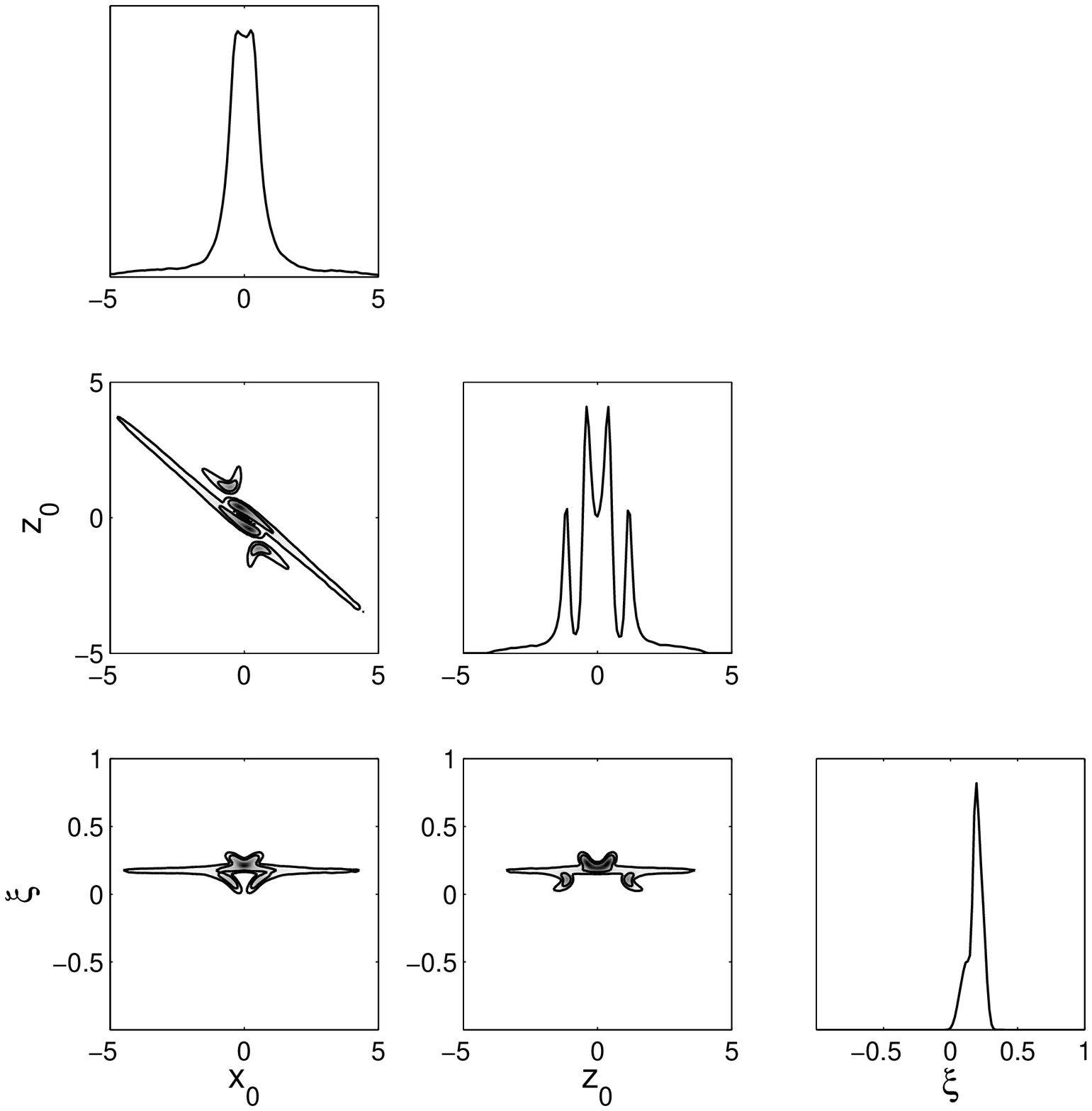}
\includegraphics[scale=0.425]{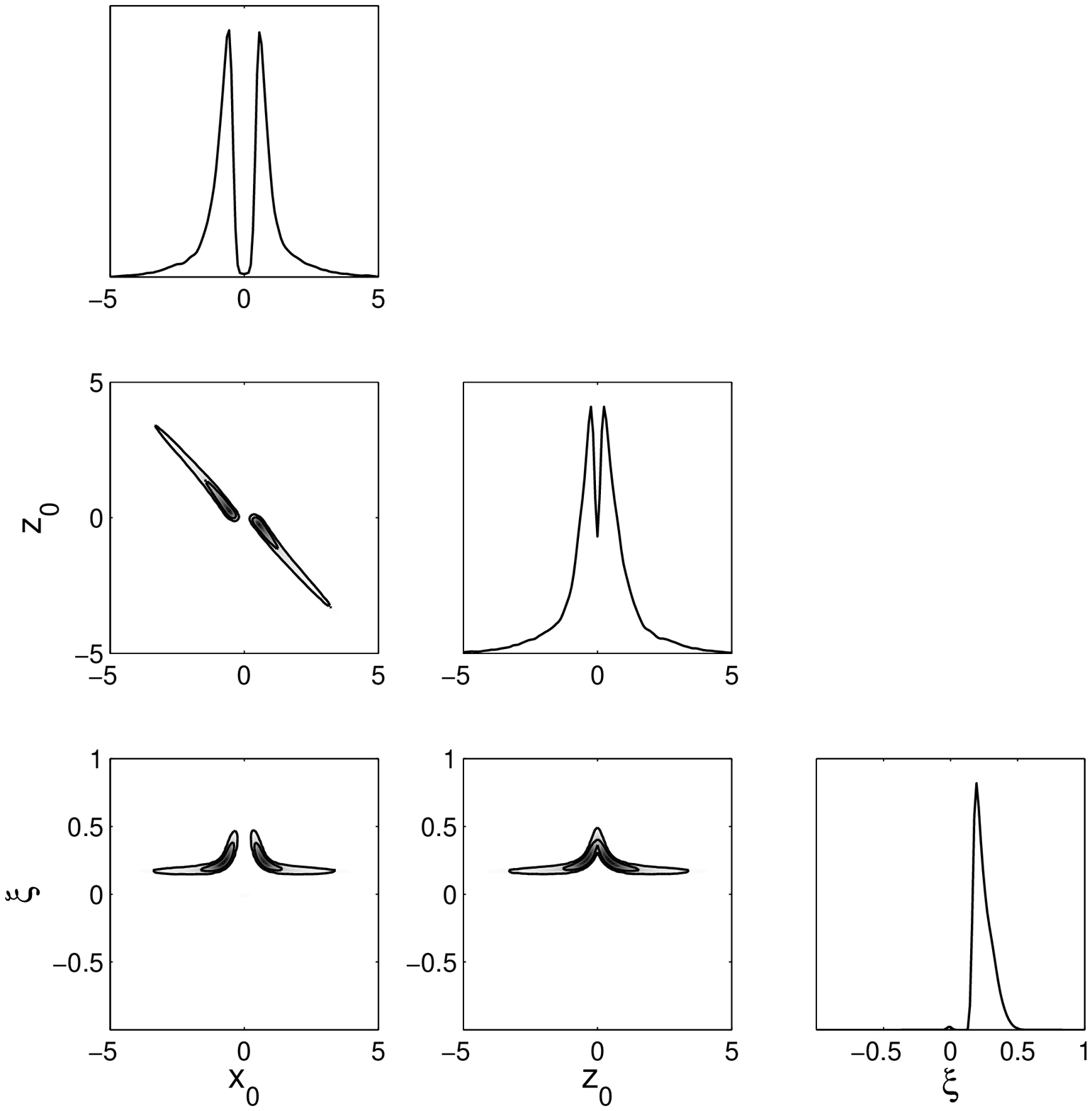}\\
\includegraphics[scale=0.425]{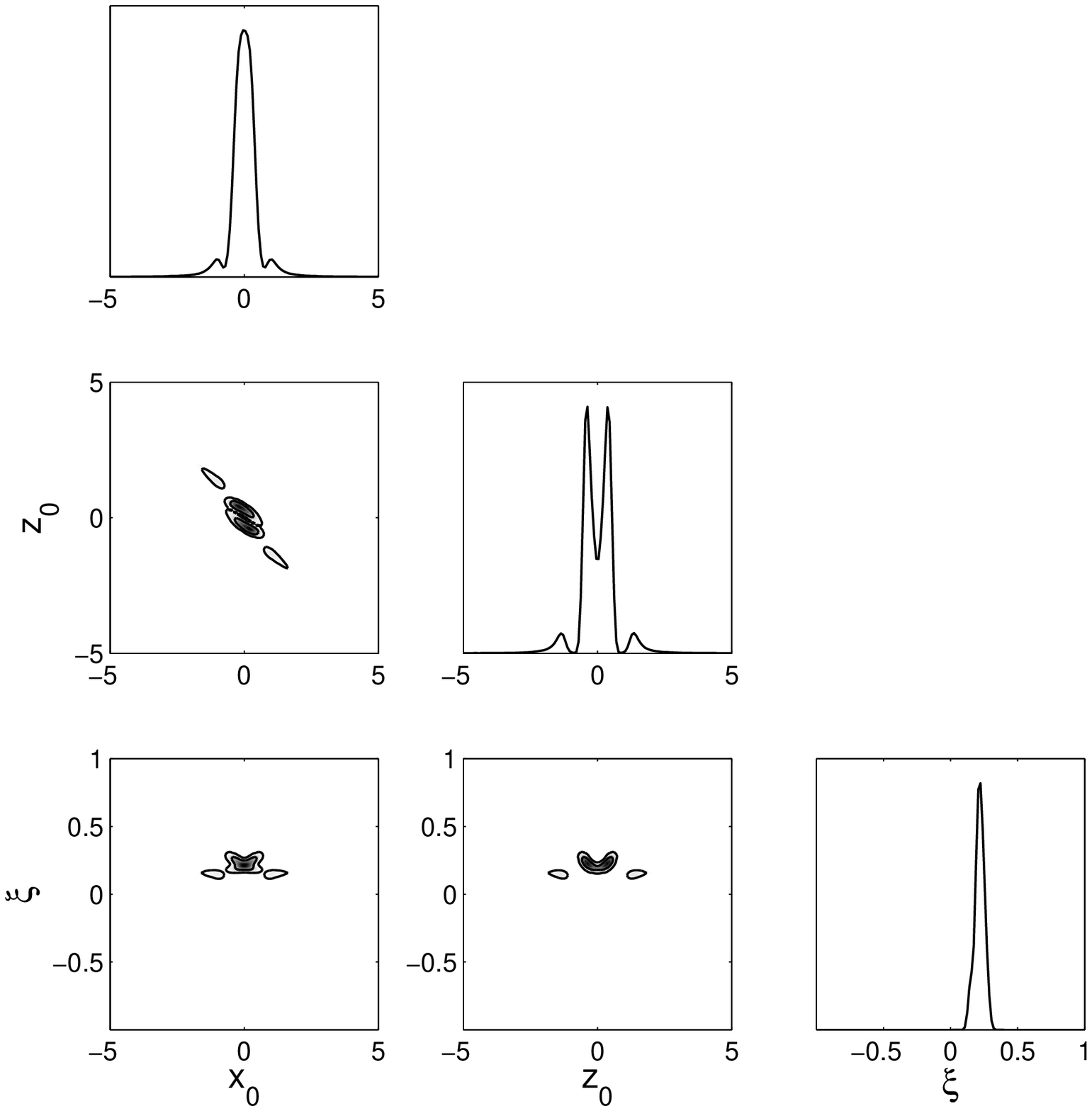}
\includegraphics[scale=0.425]{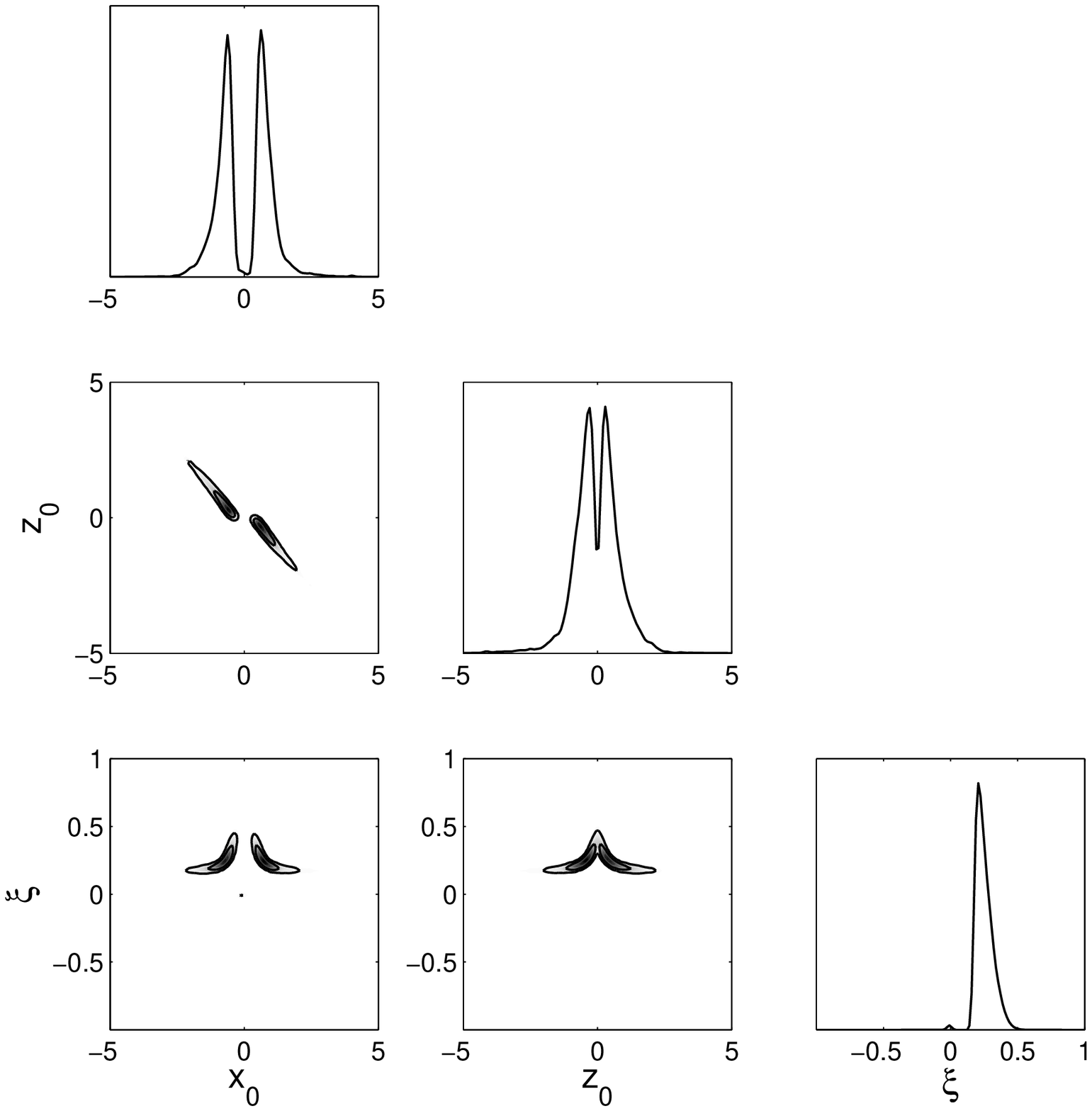}
\caption{Posterior constraints for the canonical $\ve=+1$ (left panel) and the phantom $\ve=-1$ (right panel) scalar field with the fixed baryonic matter content $\Omega_{m,0}=\Omega_{bm,0}={\rm const}$. In the bottom row the additional constraint $\Omega_{\Lambda,0}>0$ is assumed. On one-dimensional plots we present fully marginalised probabilities of the given variable, two-dimensional plots give $68\%$ and $95\%$ credible intervals of fully marginalised probabilities.}
\label{fig:1}
\end{figure*}

\begin{figure*}
\centering
\includegraphics[scale=0.425]{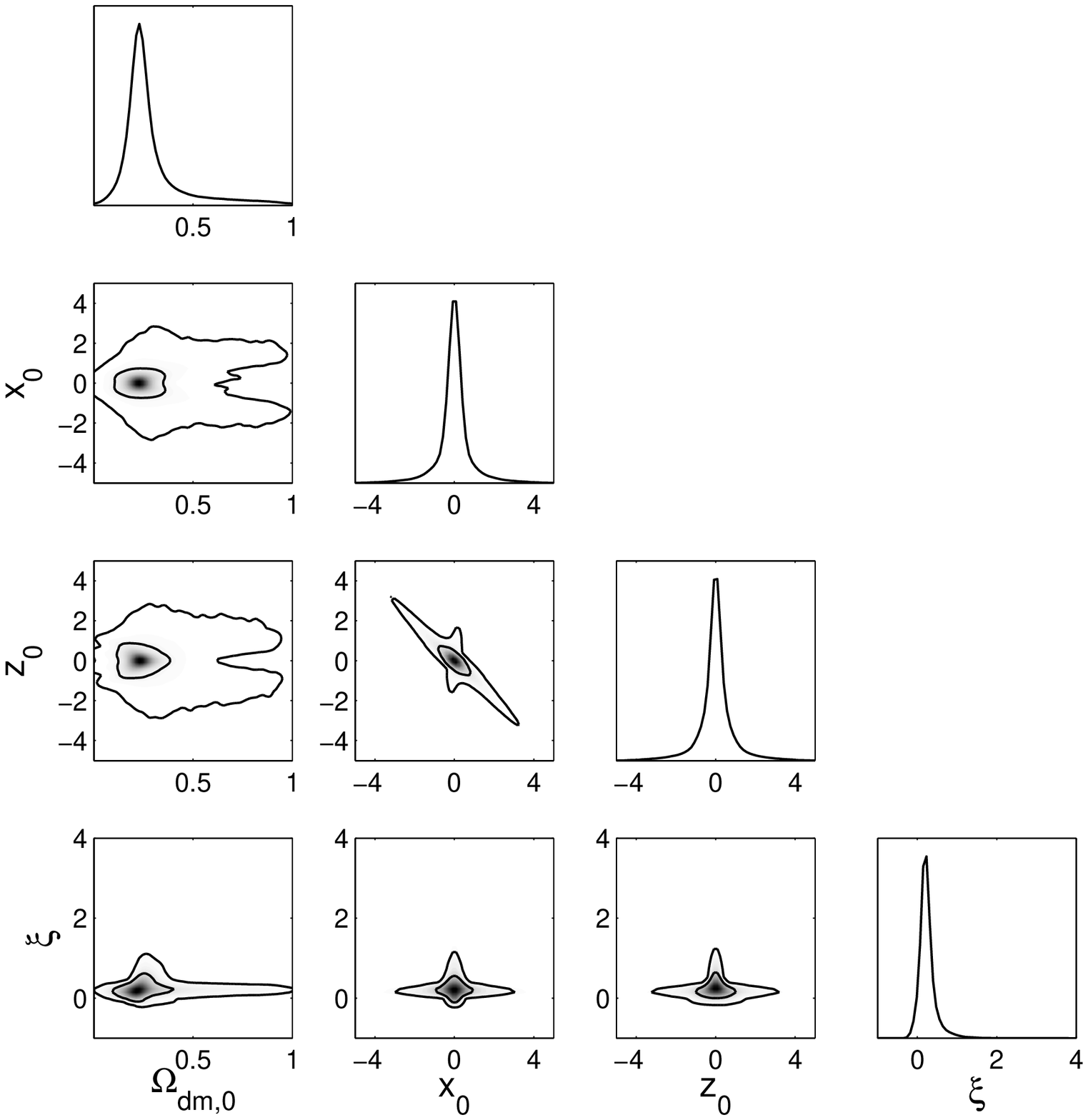}
\includegraphics[scale=0.425]{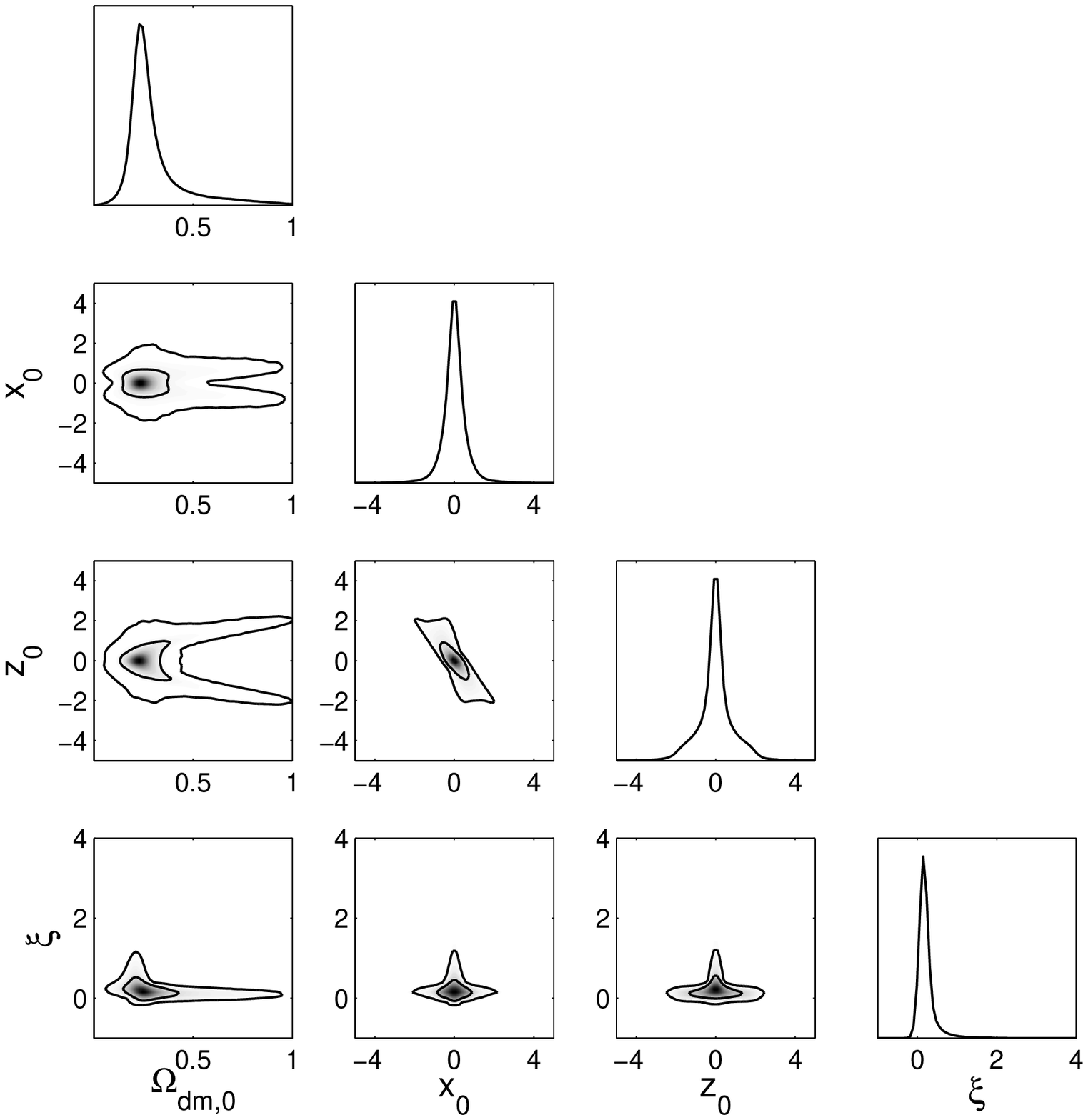}\\
\includegraphics[scale=0.425]{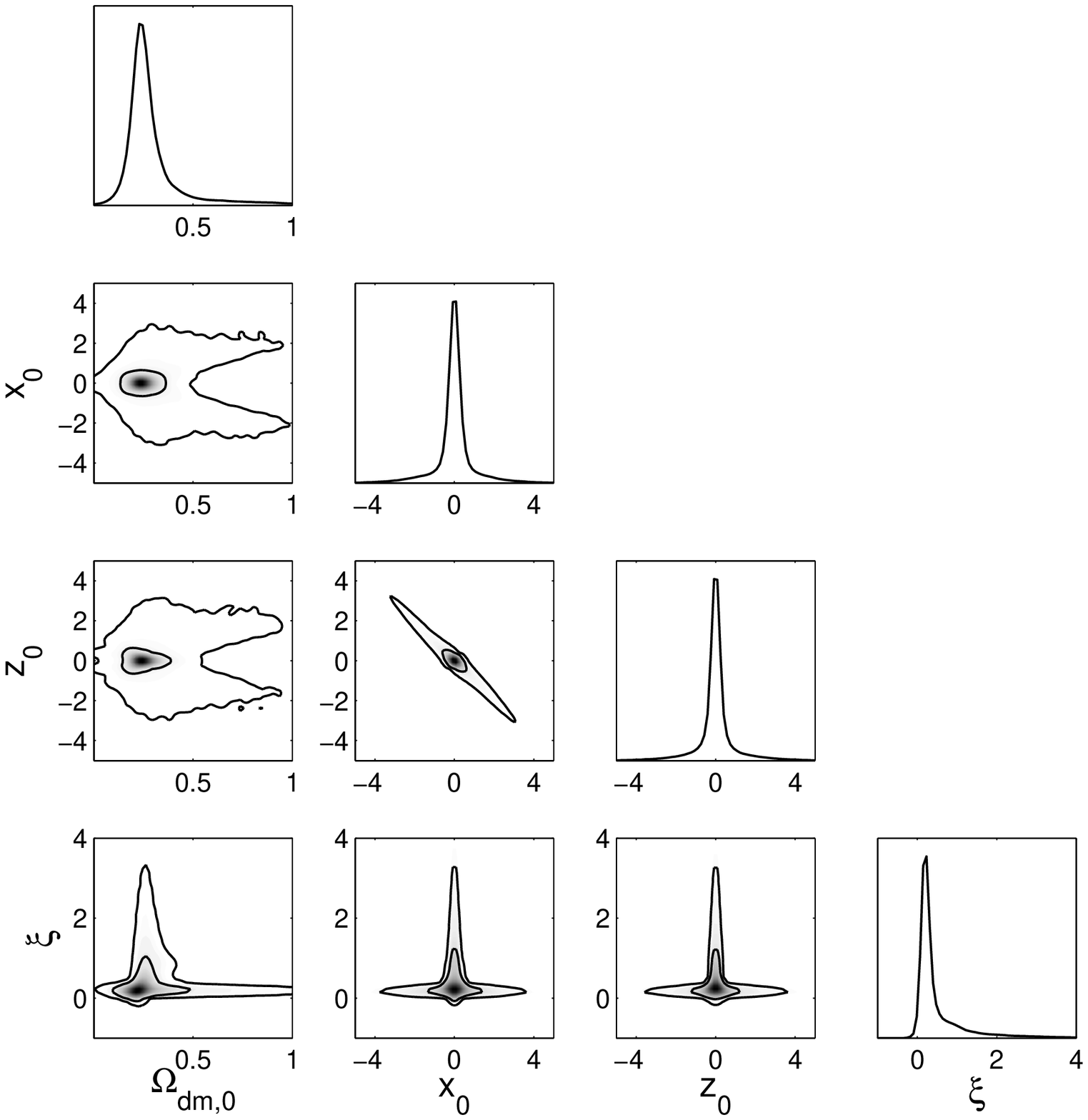}
\includegraphics[scale=0.425]{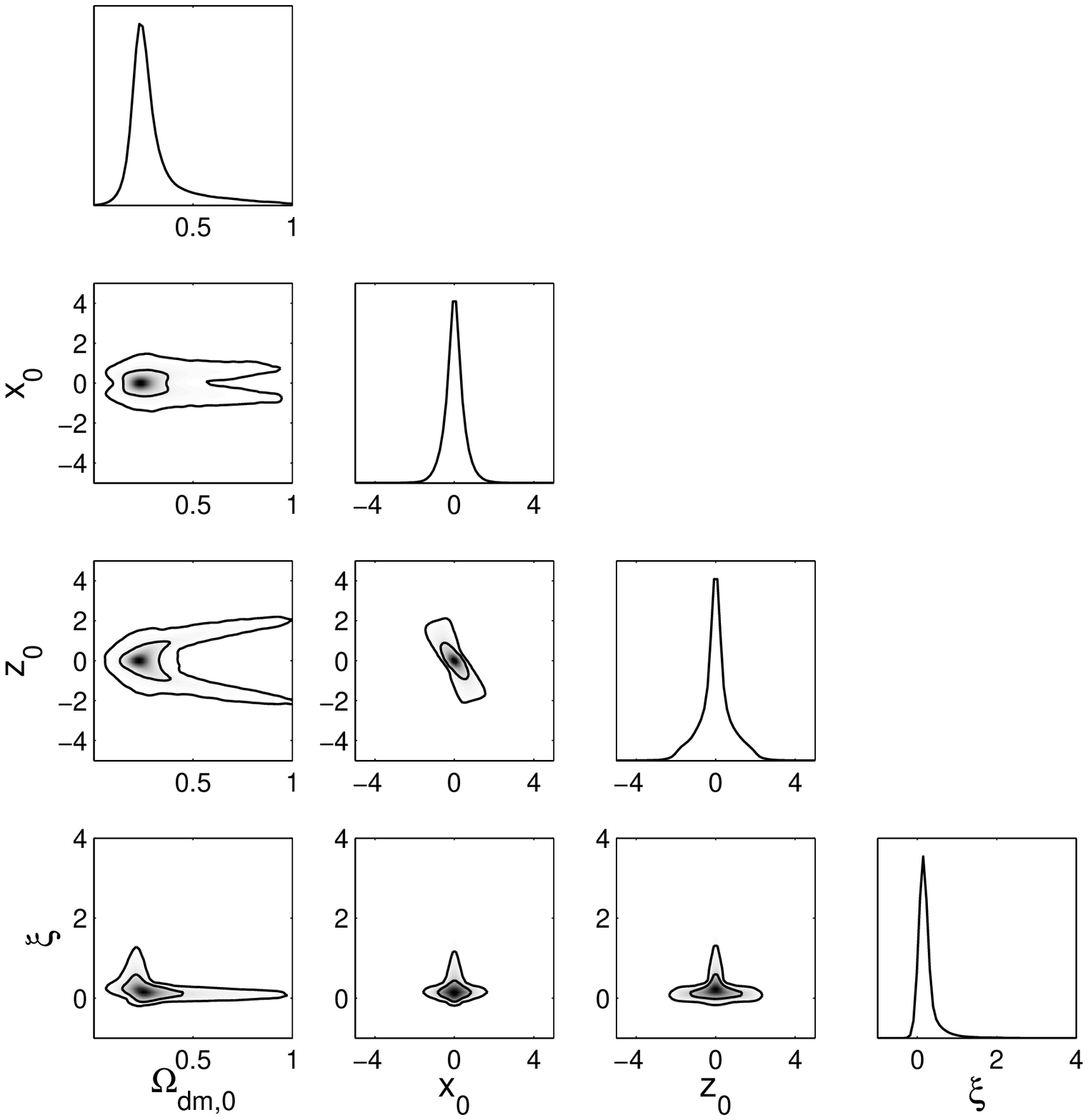}
\caption{Posterior constraints for the canonical $\ve=+1$ (left panel) and the phantom $\ve=-1$ (right panel) scalar field and the barotropic dark matter $\Omega_{m,0}=\Omega_{bm,0}+\Omega_{dm,0}$. In the bottom row the additional constraint $\Omega_{\Lambda,0}>0$ is assumed. On one-dimensional plots we present fully marginalised probabilities of the given variable, two-dimensional plots give $68\%$ and $95\%$ credible intervals of fully marginalised probabilities.}
\label{fig:2}
\end{figure*}

The $95\%$ confidence intervals for the non-minimal coupling constant for the models are the following: 
canonical 1: $\xi\in(0.0582;0.2705)$, phantom 1: $\xi\in(0.1704;0.4017)$ and additionally for the models with constraint $\Omega_{\Lambda,0}>0$, canonical 2: $\xi\in(0.1449;0.2779)$ phantom 2: $\xi\in(0.1735;0.3950)$. In the second part of table \ref{tab:1} we present models with the barotropic dark matter. The $95\%$ confidence intervals for the non-minimal coupling constant are: canonical $1+$dm: $\xi\in(-0.0371;0.8156)$, phantom $1+$dm: $\xi\in(0.0058;0.8571)$, and for models with $\Omega_{\Lambda,0}>0$, canonical $2+$dm: $\xi\in(0.0633;2.6051)$, phantom $2+$dm: $\xi\in(-0.0078;1.0186)$.

Note that the equation of motion for the scalar field with vanishing potential function non-minimally coupled to the scalar curvature 
\begin{equation}
\Box\phi-\xi R\phi=0\,,
\end{equation}
is in general not conformally invariant. However, in $n\ge2$ space-time dimensions can be made conformally invariant. Using a conformal or Weyl transformation which is the point dependent rescaling of the metric and the scalar field
\begin{equation}
\tilde{g}_{\mu\nu}=\Omega^{2}g_{\mu\nu}\,,\quad \tilde{\phi}=\Omega^{-\frac{n-2}{2}}\phi\,,
\end{equation}
where $\Omega(x)$ is a regular, nowhere vanishing function on a smooth manifold we obtain
\begin{equation}
\tilde{\Box}\tilde{\phi}-\xi \tilde{R}\tilde{\phi}=\Omega^{-\frac{n+2}{2}}\Big(\Box\phi-\xi R\phi\Big)=0\,,
\end{equation}
and this equation holds iff the non-minimal coupling constant is 
\begin{equation}
\xi=\xi_{\textrm{conf}}=\frac{1}{4}\frac{n-2}{n-1}\,.
\end{equation}
In this way we obtain a discrete set of theoretically allowed values of the non-minimal coupling constant suggested by the conformal invariance condition of the scalar field in $n\ge2$ space-time dimensions:
\begin{equation}
\begin{split} 
& \big\{(n,\xi)\big\}= \\ & \left\{\left(2,0\right),\left(3,\frac{1}{8}\right),\left(4,\frac{1}{6}\right),\left(5,\frac{3}{16}\right),\left(6,\frac{1}{5}\right),\dots,\left(\infty,\frac{1}{4}\right)\right\}\,.
\end{split}
\end{equation}
We observe that those theoretically motivated values of the non-minimal coupling constant are contained in the intervals obtained using the observational data.

The posterior constraints for all investigated models are presented in figures \ref{fig:1} and \ref{fig:2}. The one-dimensional plots we give fully marginalised probabilities of the given parameter while on the two-dimensional plots we show $68\%$ and $95\%$ credible intervals of fully marginalised probabilities. For all investigated models the fully marginalised probability distribution functions of the non-minimal coupling constant $\xi$ take the largest value in the vicinity of $\xi\approx0.2$. In the case of models without dark matter presented in figure \ref{fig:1}, the fully marginalised probability distribution functions of the non-minimal coupling constant $\xi$ are very narrow and concentrated, while the PDFs for the initial conditions of the phase space variables $(x_{0},z_{0})$ have multimodal distributions. This is due to the fact that the dynamical equations governing the background evolution of the universe have to mimic effectively dark matter and the dynamics crucially depends on the initial conditions as well as the value of the non-minimal coupling constant which are estimated form the observational data. Within the approach presented in this paper we can observe that the numerically calculated Hubble function for a given set of parameters can be expanded in to the following Taylor series 
\begin{equation}
\begin{split}
\left(\frac{H(a)}{H(a_{0})}\right)^{2} & = h^{2}(a,\Omega_{bm,0},\ve,\xi,x_{0},z_{0})\\
& \hspace{-1.75cm}\approx \Omega_{\Lambda,0}+\Omega_{1}\left(\frac{a}{a_{0}}\right)^{-1} + \Omega_{2}\left(\frac{a}{a_{0}}\right)^{-2} + \Omega_{3}\left(\frac{a}{a_{0}}\right)^{-3} + \dots
\end{split}
\end{equation}
where the density parameters are $\Omega_{i}=\Omega_{i}(\Omega_{bm,0},\ve,\xi,x_{0},z_{0})$. Using the energy conservation condition \eqref{eq:constr} we have
\begin{equation}
\Omega_{1}+\Omega_{2}+\Omega_{3}+\dots = \Omega_{bm,0}+\ve(1-6\xi)x_{0}^{2}+\ve6\xi(x_{0}+z_{0})^{2}\,.
\end{equation}
With the observational data used in this paper we can assume that $\Omega_{i}\approx0$ for $i>3$. Additionally, the $\Lambda$CDM model is favoured by the data and we can expect that $\Omega_{1}\approx0$ and $\Omega_{2}\approx0$. Thus we obtain that the leading term in the Taylor series above is the following
\begin{equation}
\Omega_{3} \approx \Omega_{bm,0}+\ve(1-6\xi)x_{0}^{2}+\ve6\xi(x_{0}+z_{0})^{2}\,.
\end{equation}
Finally, the last terms in this formula can be interpreted as an effective dark matter in the model 
\begin{equation}
\Omega_{dm,0}=\ve(x_{0}^{2}+12\xi x_{0}z_{0}+6\xi z_{0}^{2})\,,
\end{equation}
resulting from the present evolution of the scalar field. One can easily find regions in the space of parameters $(\xi,x_{0},z_{0})$ where this quantity is positive for the canonical $\ve=+1$ and the phantom $\ve=-1$ scalar field.

For the models with substantial form of dark matter presented in figure \ref{fig:2} the fully marginalised PDFs of the phase space initial conditions have unimodal distributions with sharp peak in the vicinity of zero. In the previous section we indicated that for $x_{0}\approx z_{0} \approx 0$ from the acceleration equation \eqref{eq:accel_int} one obtains 
\begin{equation}
\left(\frac{H(a)}{H(a_{0})}\right)^{2}=\Omega_{\Lambda,0}+\Omega_{m,0}\left(\frac{a}{a_{0}}\right)^{-3}+\mathcal{O}(\xi,x_{0},z_{0})\,,
\end{equation}
where now $\Omega_{m,0}=\Omega_{bm,0}+\Omega_{dm,0}$ and the dark matter contribution is one of the estimated parameters. We observe that the terms dependent on the present values of phase space variables constitute small deviation from the standard $\Lambda$CDM model.

\begin{table}
\renewcommand{\arraystretch}{1.5}
	\centering
	\begin{tabular}{lcc}
		\hline
model & evidence $\ln{E_{i}}$ & $2\ln{B_{0i}}$\\
		\hline
canonical $1$	& $-293.53\pm0.13$ &  $16.88\pm0.42$ \\
phantom $1$	    & $-294.67\pm0.37$ &  $19.17\pm0.76$ \\
canonical $2$	& $-293.68\pm0.23$ &  $17.19\pm0.49$ \\
phantom $2$	    & $-294.34\pm0.31$ &  $18.50\pm0.65$ \\
		\hline
canonical $1+$dm  & $-292.16\pm0.21$ & $14.15\pm0.45$ \\
phantom $1+$dm    & $-292.66\pm0.31$ & $15.14\pm0.65$ \\
canonical $2+$dm  & $-291.77\pm0.18$ & $13.36\pm0.41$\\
phantom $2+$dm	  & $-292.28\pm0.24$ & $14.39\pm0.51$\\
\hline
$\Lambda$CDM & $-285.09\pm0.10$& $0$ \\
\hline
	\end{tabular}
	\caption{Values of the evidence and the Bayes factor with respect to the $\Lambda$CDM model.}
\label{tab:2}
\end{table}

In order to discriminate between models we used twice of the natural logarithm of the Bayes factor of two models defined as 
\begin{equation}
2\ln{B_{0i}}=2\ln{\frac{E_{0}}{E_{i}}}\,,
\end{equation}
which is proportional to the ratio of the evidence of the base model $E_{0}$ and the evidence of the investigated model $E_{i}$.

The value of the Bayes factor can be interpreted as a evidence in favour of the base model. For $2>2\ln{B_{0i}}$ the evidence is not worth a bare mention, for $6>2\ln{B_{0i}}>2$ is positive, for $10>2\ln{B_{0i}}>6$ is strong and, finally, the evidence in favour of the base model is very strong when $2\ln{B_{0i}}>10$ (or, equivalently, very strong evidence against the investigated model) \cite{Kass:1995}.

In table \ref{tab:2} we gathered the values of twice the natural logarithm of the Bayes factor of investigated models with respect to the $\Lambda$CDM model. The performed analysis using \emph{Union2.1+H(z)+\\Alcock-Paczy\'{n}ski} data set indicates on a very strong evidence in favour of the $\Lambda$CDM model over the investigated models.

Among the investigated models the best one is with the canonical scalar field, the substantial dark matter and the constraint $\Omega_{\Lambda,0}>0$. Nevertheless analysis based on the Bayes factor can not distinguish between different investigated models with additional contribution in the form of the dark matter since its value with the canonical $2+$dm model as the base model is less that 2. The same situation takes place for the models without dark matter, all the investigated models are indistinguishable. In the case of models with and without dark matter contribution we obtain a positive evidence for models with the substantial dark matter.

\section{Conclusions}
 

In this paper we obtained cosmological constraints on the simple cosmological models with the flat FRW symmetry filled with non-minimally coupled scalar field and a constant potential function. Using the de Sitter state invariant manifold we were able to reduce the background dynamics and we found the exact solutions for the reduced dynamics. We have shown that the investigated models constitute extensions of the standard $\Lambda$CDM model beyond the minimally $\xi=0$ and conformally $\xi=\frac{1}{6}$ coupled scalar field cases. The performed analysis using \emph{Union2.1+H(z)+Alcock-Paczy\'{n}ski} data set enable us to exclude negative values of the non-minimal coupling constant on the $68\%$ confidence level. For two investigated models with the phantom scalar field and with the fixed baryonic matter content the value of conformal coupling $\xi=\frac{1}{6}$ lies below the $95\%$ confidence interval.

We have shown that the non-minimal coupling constant $\xi$ can be a useful parameter in description of the current accelerated expansion of the universe in the quintessence domination era. The special bifurcation value of the non-minimal coupling $\xi=\frac{3}{16}$ is distinguished by the dynamics of the model, constituting the border between an oscillatory and a linear behaviour on the invariant de Sitter manifold, can lead to a singularity free cosmological evolution \cite{Hrycyna:2015eta,Hrycyna:2015b}. Finally, the performed statistical analysis may suggest some additional symmetry in the matter sector of the theory \cite{Nakayama:2013is,Bars:2013yba,Englert:1975wj,Englert:1976ep,tHooft:2011aa,tHooft:2014daa,tHooft:2015}.

\section*{Acknowledgements}
I would like to thank Marek Szyd{\l}owski for valuable discussions 
and comments. I am grateful to organisers of the XIIth School of Cosmology, September 15 - 20, 2014 held at Institut d'Etudes Scientifiques de Carg{\`e}se, for invitation and opportunity to present a part of this work and to Kei-ichi Maeda for comments. 

This research was funded by the 
National Science Centre through the postdoctoral internship award 
(Decision No.~DEC-2012/04/S/ST9/00020) and partially supported by the National Science Centre through the OPUS 5 funding scheme (Decision No.~DEC-2013/09/B/ST2/03455). The use of the 
{\'S}wierk Computing Centre (CI{\'S}) computer cluster at the 
National Centre for Nuclear Research is gratefully acknowledged.

\bibliographystyle{elsarticle-num}
\bibliography{../confinv,../numerics,../darkenergy,../quintessence,../quartessence,../astro,../dynamics,../standard,../inflation,../sm_nmc,../singularities,../moje}

\end{document}